\documentclass[twocolumn,showpacs,preprintnumbers,amsmath,amssymb]{revtex4}

\usepackage{graphicx}
\usepackage{dcolumn}
\usepackage{bm}

\begin{document}

\preprint{APS/123-QED}

\title{Interplay between Co-3d and Ce-4f magnetism in CeCoAsO}
\author{Rajib Sarkar}
\author{Anton Jesche}
\author{Cornelius Krellner}
\author{Michael Baenitz}
\author{Cristoph Geibel}
\affiliation{
Max-Planck Institute for Chemical Physics of Solids, 01187
Dresden, Germany
}%
\author{Chandan Mazumdar}
\author{Asok Poddar}
\affiliation{
ECMP Division, Saha Institute of Nuclear Physics, 1/AF Bidhannagar, Kolkata-700064, India }%
\date{\today}
\begin{abstract}
We have investigated the ground state properties of
polycrystalline CeCoAsO by means of magnetization, specific heat
and solid state NMR. Susceptibility and specific-heat measurements
suggest a ferromagnetic order at about, $T_\mathrm{C}$=75 K. No
further transitions are found down to 2 K. At 6.5 K a complex
Schottky type of anomaly shows up in the specific heat results.
The interplay between Ce-4f and Co-3d magnetism being responsible
for that anomaly is discussed. Furthermore $^{75}$As NMR
investigations have been performed to probe the magnetism on a
microscopic scale. As-NMR spectra are analysed in terms of first
and second order quadrupolar interaction. The anisotropic shift
component $K_{\mathrm{ab}}$ and $K_{\mathrm{c}}$ could be derived
from the $^{75}$As powder spectra. Towards lower temperature a
strong shift anisotropy was found. Nonetheless $K_{\mathrm{iso}}$
tracks the bulk susceptibility down to $T=$50 K very well.
Furthermore the presence of weak correlations among the Ce ions
in the ferromagnetic state is discussed. The observed increase of
$C/T$ towards lower temperatures supports this interpretation.

\end{abstract}
\pacs{Valid PACS appear here}
\maketitle
\section{\label{sec:level1}Introduction:}
The rare earth transition metal pnictides RTPnO (R: rare earth,
T: transition metal, Pn: P or As) attracted considerable attention
because of the recent discovery of superconductivity with a
transition temperatures $T_{\mathrm{C}}$ up to 50 K in the
RFeAsO$_{\mathrm{1-x}}$F$_{\mathrm{x}}$ series of compounds being
the highest $T_{\mathrm{C}}$$`$s except for cuprate systems
\cite{{Kamihara-2008},{Chen-2008},{Z. A.
Ren-2008},{Chen-Li-2008},{J. Yang-Supercond.Sci.
Technol.-2008},{J. G. Bos- Chem. Commun.-2008}}. While the main
studies on these materials are devoted to the superconducting
materials, the non superconducting members of these family stays
mostly unexplored. Nevertheless studying these compounds may
provide information to understand also the superconducting state.
Recent studies of CeRuPO and CeOsPO
\cite{{Krellner-2007},{Krellner-Crystal frowth-2007},{C.
Krellner-prl-2007}} indicate dissimilar types of magnetic
ordering. CeRuPO is a rare example for ferromagnetic (FM) Kondo
system showing a FM order at $T_{\mathrm{C}}$=15 K and a Kondo
energy scale of about $T_\mathrm{K}$ $\simeq$ 10 K. CeOsPO
exhibits an antiferromagnetic (AFM) order at $T_{\mathrm{N}}$=4.5
K.  The recent studies of the relative CeFePO suggest that this
is a heavy Fermion metal  with strong correlation of the
4f-electrons close to a magnetic instability \cite{E. M.
Bruning-prl-2008}. On the other hand in CeFeAsO, a complex
interplay of Ce-4f and Fe-3d magnetism is found. Here, Ce orders
antiferromagnetically at $T_{\mathrm{N}}$$\simeq$3.8 K whereas the
high temperature regime is dominated by the 3d magnetism of
Fe\cite{Jun Zhao-nature
materials-2008}\cite{Anton-CeFeAsO-NJP-2009}. There is a
structural transition from tetragonal to orthorhombic at $\simeq$
151 K followed by SDW type AFM order of Fe at $\sim$ 145 K.
Moreover Ce magnetism is not effected strongly by the presence of
the Fe moments. Furthermore, neutron scattering, muon spin
relaxation experiments and recent analysis of $\chi(T)$, $C(T)$
suggest that there is a sizeable inter-layer coupling in CeFeAsO
\cite{{Anton-CeFeAsO-NJP-2009}, {Anton-CeFeAsO-to be
published-2010}, {Chi S-prl-2008},{Maeter condmat-2009}}.
Therefore, the pure CeFeAsO system have already proven to be a
rich reservior of exotic phenomena.
\\
\indent Apart from isoelectronic substitution on CeTPnO with Fe,
Ru, Os chemically also the T= Co series form.  Results on LaCoAsO
and LaCoPO were reported to exhibit ferromagnetic order of Co
moment with Curie temperatures $T_{\mathrm{C}}$ of about
$T_{\mathrm{C}}$=50 K and $T_{\mathrm{C}}$=60 K, respectively. In
contrast to Fe, where the 3d magnetism depends on the
Pnictide(P,As), Co stays magnetic in both the P and As series. In
LaCoAsO, Co saturation moments of 0.3-0.5 $\mu_\mathrm{B}$ per Co
\cite{{H.Hosono-LaCoAsO-PRB-2008},{Mandrus-LaCoAsO-PRB-2008},{H.
Ohta and K. Yoshimura-Phys rev. B-2009}} are found.
It has been proposed that spin fluctuations play an important
role in the magnetic behavior of LaCoAsO
\cite{H.Hosono-LaCoAsO-PRB-2008}\cite{H. Ohta and K.
Yoshimura-Phys rev. B-2009}, as well as the magnetic and
superconducting properties of the iron-based superconductors.
Last year we reported the detailed physical properties of CeCoPO
and discussed about the interplay between 3d magnetic moments of
Co and 4f electrons of Ce \cite{C. Krellner-Physica B-2009}. This
system, similar to LaCoPO, the Co-3d electrons order
ferromagnetically. However, here the Ce-ions are on the border to
magnetic order and an enhanced Sommerfeld coefficient, $\gamma
\sim$ 200 mJ/mol K$^{2}$ was found. In CeTPnO the substitution of
P by As change the magnetism drastically. It is already evident in
the case of CeFeAsO \cite{{Yongkang Luo-condmat-2009},{Clarina de
la Cruz-condmat-CeFeAs1-xPxO-2009},{Rajib
Sarkar-CeFeAs1-xPxO-2010}}. Therefore it is natural to investigate
the physical and microscopic properties of CeCoAsO compound.
\\
In this report we present the physical properties of
polycrystalline  CeCoAsO using susceptibilty $\chi(T)$ and
specific heat $C(T)$ measurements. Additionally, we discuss the
preliminary microscopic results as seen by $^{75}$As NMR study.
\section{Experimental}
Samples are prepared by solid state reaction technique. The
starting materials that are taken for the preparation of parent
CeCoAsO are Ce and As chips and Co, Co$_{3}$O$_{4}$ powders.
First, CeAs was prepared by taking stoichiometric amounts of Ce
and As in 1:1 ratio, pressed into pellets and sealed in evacuated
quartz tube. With repeated heat treatment, attaining a maximum
temperature of 900$^{\circ}$C, and grinding inside a glove box
filled with inert Ar gas, single phase CeAs were obtained. CeAs
were then mixed thoroughly with Co$_{3}$O$_{4}$ and Co powder in
stoichiometry and pressed into pellets. The pellets were wrapped
with Ta foil and sealed in an evacuated quartz tube. They were
then annealed at 1100-1150$^{\circ}$C for 40-45 hours to obtain
the final CeCoAsO samples. X-ray powder diffraction revealed a
single phase sample with no foreign phases. Susceptibility
$\chi(T)$ measurements were performed in a commercial Quantum
Design (QD) magnetic property measurement system (MPMS). Specific
heat $C(T)$, measurements were performed in a QD physical
property measurement system (PPMS). For the NMR measurements,
polycrystalline powder was fixed in the paraffin to ensure a
random orientation. $^{75}$As NMR measurements were performed
with a standard pulsed NMR spectrometer (Tecmag) at the frequency
48 MHz as a function of temperature. The field sweep NMR spectra
were obtained by integrating the echo in the time domain and
plotting the resulting intensity as a function of the field.
Shift values are calculated from the resonance field $H^{\ast}$
by $K(T)=(H_{\mathrm{L}}-H^{\ast})/H^{\ast}$ whereas the Larmor
field, $H_{\mathrm{L}}$, is given by using GaAs ($^{75}$As-NMR)
as reference compound with $^{75}K\simeq0$ \cite{Basto-GaAs-1999}.
\section{Results}
\subsection{\label{sec:level2}Magnetisation and specific heat study}
Fig. {\ref{fig:susceptibilty} shows the susceptibility of CeCoAsO
as a function of  temperature at different fields as indicated.
Above 200 K, the susceptibility follows the Curie Weiss behavior
with an effective moment $\mu_{\mathrm{eff}}$= 2.74
$\mu_{\mathrm{B}}$. This value is higher than the value of
$\mu_{\mathrm{eff}}^{Ce}$=2.54 $\mu_{\mathrm{B}}$, expected for a
free Ce$^{3+}$ ion. This is because of the contribution of Co to
the effective moment of
$\mu_{\mathrm{eff}}=\sqrt{\mu^{(\mathrm{Co}) 2}
+\mu^{(\mathrm{Ce}) 2}}$. For the Co ions we calculate an
effective moment of $\mu_{\mathrm{eff}}^{Co}$=1.03
$\mu_{\mathrm{B}}$. Our results are in good agreement with
findings from Ohta and Yoshimura et. al. \cite{H. Ohta and K.
Yoshimura-Phys rev. B-2009}. A sharp increase of the
susceptibility is observed at around 75 K. Here, a strong field
dependence of the susceptibility typical for a FM system is
evidenced. In the inset of  Fig. {\ref{fig:susceptibilty} the
magnetisation M(H) of CeCoAsO up to 5 T at 2 K is shown. A large
hysteresis typical for hard ferromagnets is observed. A large
hysteresis was also found for CeCoPO whereas for other RCoAsO no
such large hysteresis was found \cite{H. Ohta and K.
Yoshimura-Phys rev. B-2009}.
\begin{figure}
\includegraphics[scale=0.90]{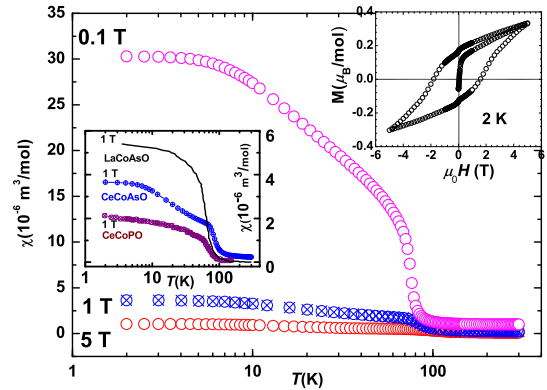}
\caption{\label{fig:susceptibilty} Temperature dependence of the
susceptibility of CeCoAsO at different fields. The right inset
shows the magnetisation  at 2 K and the left inset shows a
comparison of the susceptibility for CeCoPO (data taken from
\cite{C. Krellner-Physica B-2009}), CeCoAsO, and LaCoAsO (data
taken from \cite{Athena S. Sefat-LaCoAsO-PRB-2008}) at 1 T. }
\end{figure}
\indent In the left inset of Fig. {\ref{fig:susceptibilty} we
compare the susceptibility results of CeCoAsO with that of
LaCoAsO and CeCoPO at 1 T. It should be mentioned that for CeCoAsO
at smaller fields no peak could be resolved which is in contrast
to CeCoPO \cite{C. Krellner-Physica B-2009}. The temperature
dependence of the susceptibilty is quiet similar for CeCoAsO and
CeCoPO, whereas there is a pronounced difference to that of
LaCoAsO.
In LaCoAsO there is no influence of 4f magnetism. Therefore the
magnetisation curve looks like a simple textbook ferromagnet. It
has to be mentioned that we took the data for LaCoAsO from Ref.
\cite{Athena S. Sefat-LaCoAsO-PRB-2008}. For this set of data
$\chi$(LaCoAsO)$>$ $\chi$(CeCoAsO) is found in the FM state but
surprisingly the data by Ohta et al. \cite{H. Ohta and K.
Yoshimura-Phys rev. B-2009} shows smaller values (at same field)
suggesting $\chi$(LaCoAsO)$<$ $\chi$(CeCoAsO). Nonetheless our
results for CeCoAsO are in perfect agreement with findings in
\cite{H. Ohta and K. Yoshimura-Phys rev. B-2009}. The different
behavior for CeCoPO and CeCoAsO indicate that the Ce-4f creates a
significant change in the overall magnetic behavior. This unusual
behavior of the susceptibility indicates an intricate magnetic
structure with strong polarisation of the itinerant Co moments on
the more localised Ce moments. Such effects are also known from
4f-ion Fe$_{4}$Sb$_{12}$ skutterudites. Here EuFe$_{4}$Sb$_{12}$
shows similar magnetisation
curves\cite{Bauer-Skutter-EuFe4Sb12-2001}. One approach could be
to describe CeCoAsO in the framework of a classical ferrimagnet
like RCo$_5$ with R=Y, Ce, Pr.... \cite{France-Radwanski-1993}.
Magnetisation in these system is goverened by the two subsystems
of the 4f and 3d moments and their inter and intra molecular
interactions. Because of the antiferromagnetic coupling of the
rare earth spins with the Co spin for light rare earth ions like
Ce, usually an ferromagnetic alignment of the Ce-4f moment and
the Co-3d moments is expected \cite{France-Radwanski-1993} in the
ordered state. This would imply $\chi$(CeCoAsO)$<$
$\chi$(LaCoAsO). Unfortunately because of inconsistent literature
for LaCoAsO data we have no prove for that. To summarize the
magnetisation section it should be note that besides the high
temperature ordering at $T_{C}$=75 K for CeCoPO and $T_{C}$=75 K
for CeCoAsO no further transitions are evident from magnetisation
measurements. This is in contrast to other RCoAsO system with
R=Gd, Sm or Nd, where the rare earth moments orders
antiferromagnetically at low temperature\cite{H. Ohta and K.
Yoshimura-Phys rev. B-2009}\cite{David
Mandrus-NdCoAsO-condmat-2009}.
\begin{figure}
\includegraphics[scale=0.90]{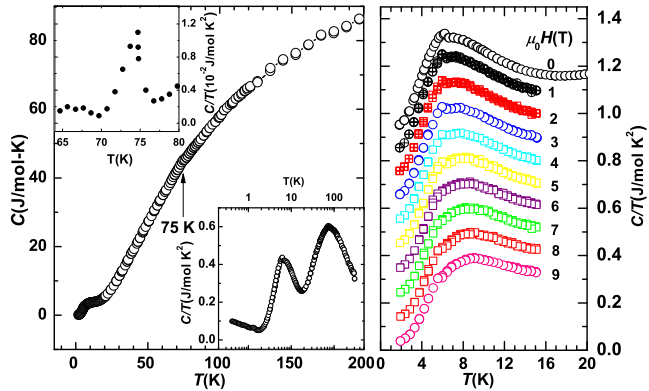}
\caption{\label{fig:specific} (Left panel)Temperature dependence
of the specific heat of CeCoAsO. The right inset shows the
specific heat plotted as $C/T$ on a logarithmic temperature scale.
The left inset shows the behavior of transition after subtracting
the background. Right panel shows the specific heat plotted as
$C/T$ vs. $T$ at different field in the temperature range 1.8 -20
K (note that 0.1 J/mol-K$^{2}$ was added as an offset in the C/T
plot on the right). }
\end{figure}
\indent The temperature and field dependence of the specific heat
$C(T)$ of CeCoAsO is shown in Fig. {\ref{fig:specific}.
Towards high temperatures $C(T)$ converges nicely to the classical
Dulong Petit limit of $\sim$ 100 J/mol-K. By lowering the
temperature a broad anomaly at $T_{\mathrm{C}}$ $\sim$75 K is
visible on the top of the phonon dominated specific heat. We have
estimated the background using the third order polynomial from 65
to 85 K, excluding the temperature range near the peak at 75 K.
Then the background was subtracted from the data to obtain
specific heat. In the left inset we have shown the $C(T)/T$ vs.
$T$ plot after subtracting the background. It is worthwhile to
mention that at the same temperature (75 K) the susceptibility
increases sharply. Therefore, this anomaly is being due to the
ferromagnetic ordering of Co. At around 6.5 K an additional broad
anomaly in the $C(T)$ shows up. The right inset shows the
$C(T)/T$ vs. $T$ plots. In this inset both anomalies are rather
pronounced. To understand the origin of the low temperature
anomaly we have investigated the field dependent specific heat in
the temperature range 1.8 - 15 K and in the field range 0-9 T. The
right panel of Fig. {\ref{fig:specific} shows the $C/T$ vs. $T$
plot at different fields (curves are shifted on y axis by 0.1
J/mol-K$^{2}$). It seems that the effect of field on $C(T)$ is
very small. Nonetheless, the broad maxima is shifted
insignificantly towards higher temperatures with increasing
field. The preliminary analysis of the specific heat reveal that
this broad anomaly at low temperature is not due to the ordering
of Ce. Rather this is reminiscent of Schottky type anomaly. This
might be attributed to the level splitting CEF ground state of Ce
by the polarisation field of Co. For Ce$^{3+}$ ion is, in a
tetragonal environment, $\mathrm{J}=5/2$ splits into three
doublets. No further degeneracy can be removed by the CEF.
However, these levels can be further splitted by the exchange
field of the moments and/or the internal field of the 3d and 4f
moments.\cite{Anton-CeFeAsO-to be published-2010}\cite{Chi
S-prl-2008}\cite{{Cheng-NdMnO3-2005},{Hagary-RCo9Si4-2008},{Lima-2002},{Aeby-CeP-CeAs-1973},{Zapf-2003}}
\begin{figure}
\includegraphics[scale=0.99]{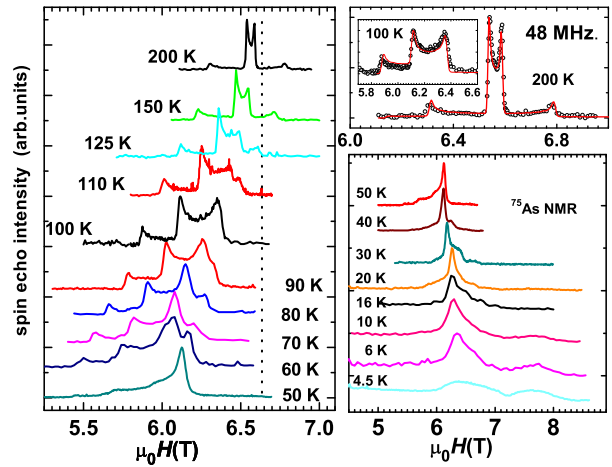}
\caption{\label{fig:Spectra} Temperature dependence of the
$^{75}$As field sweep NMR spectra at 48 MHz down to 50 K (left
panel). Dotted line indicates the Larmor field estimated from the
reference compound GaAs with $^{75}K\simeq0$
\cite{Basto-GaAs-1999}. $^{75}$As spectra below 50 K (right,
bottom). $^{75}$As spectra at 200 K and 100 K together with the
theoretical simulation (solid line)(right, top).}
\end{figure}
\\
\indent Below 1.8 K, $C/T$ increases logarithmically and tends to
saturate at further low temperature at $\gamma \sim$ 200
mJ/mol-K$^{2}$. This enhancement might indicate the presence of
strong correlations between the Ce ions in the Co dominated
ferromagnetic ground state.
\\
\indent Furthermore, at zero field we have estimated the entropy
gain by integrating the 4f part to the specific heat
$C_{\mathrm{4f}}/T$ in the temperature range 1.8-15 K. For the
estimation of the entropy gain we have subtracted the phonon
contribution by using reference compound data of LaCoAsO in the
temperature range 1.8-15 K after Sefat et. al. \cite{Athena S.
Sefat-LaCoAsO-PRB-2008}. However, the contribution of LaCoAsO to
the specific heat is small below 15 K. The estimated entropy gain
for CeCoAsO at 15 K is 75$\%$ of \textit{R}ln2. This supports the
scenario based on the splitting of the CEF doublet ground state.
\subsection{\label{sec:level2}$^{75}$As NMR}
 Fig.{\ref{fig:Spectra} shows the field sweep $^{75}$As NMR spectra at
different temperatures. Because $^{75}$As is a I=3/2 nuclei, the
quadrupole interaction should be taken into account for the
interpretation of the spectra. The main effects are I.$)$ first
order interaction, occurrence of pronounced satellite peaks
(3/2$\leftrightarrow$1/2, -1/2$\leftrightarrow$-3/2), II.$)$
second order interactions, splitting of the central
-1/2$\leftrightarrow$1/2 transition.  At high temperature the
second order quadrupolar splitting central transition along with
the two first order satellite transitions are observed and the
spectra could be nicely simulated (Fig.{\ref{fig:Spectra}, right,
top). Initially, with lowering of temperature down to 50 K, the
whole spectra is shifted towards the low field side with
considerable line broadening and develops large anisotropy.
However, below 50 K the whole spectra shifts towards the high
field side with further gradual line broadening. The line
broadening of the spectra become enormous below 15 K. It is clear
from  Fig. {\ref{fig:Spectra} that the NMR spectra become more
and more complex by lowering the temperature and the simulation
of this broad spectra, at low temperature, is rather complicated.
Nevertheless, with considerable effort it is possible to identify
the singularities of these spectra down to 20 K. Therefore,
estimation of the shift is possible by fitting the singularity of
the spectra down to 20 K. During the compilation of this paper we
realize that similar analysis has been performed for LaCoAsO
\cite{H. Ohta and K. Yoshimura-Condmat-LaCoAsO-NMR-2009}. In Fig.
{\ref{fig:Spectra} (Lower right panel) below 10 K a considerable
background intensity is perceived in the low field side. This
background signal is the combination of the $^{59}$Co spectra and
the $^{75}$As NMR spectra. We already measured some $^{59}$Co NMR
spectra. The estimated $\nu_{\mathrm{Q}}$ value from $^{75}$As NMR
spectra at high temperature is 3.6 MHz. This value is similar to
that of LaCoAsO system \cite{H. Ohta and K.
Yoshimura-Condmat-LaCoAsO-NMR-2009} and somewhat smaller than
what was found for the CeFeAsO system \cite{Rajib
Sarkar-CeFeAs1-xPxO-2010}. While lowering the temperature
$\nu_{\mathrm{Q}}$ monotonically increases and down to 20 K no
drastic changes could be detected. This rules out a sudden
structural change in this compound. From the simulation of the
spectra, we have estimated the shift components
$^{75}$K$_{\mathrm{ab}}$ and $^{75}$K$_{\mathrm{c}}$
corresponding to $H\bot \mathrm{c}$ and $H\parallel \mathrm{c}$
direction, respectively.
\\
\indent Fig. {\ref{fig:Shift} shows the variation of
$^{75}$K$_{\mathrm{ab}}$, $^{75}$K$_{\mathrm{c}}$ and
$^{75}$K$_{\mathrm{iso}}$ as a function of temperature.
$^{75}$K$_{\mathrm{iso}}$ was estimated using the equation
$^{75}K_{\mathrm{iso}} = \frac{2}{3}^{75}K_{\mathrm{ab}}
+\frac{1}{3} ^{75}K_{\mathrm{c}}$. From  Fig. {\ref{fig:Shift} it
is evident that $^{75}$K$_{\mathrm{ab}}$ and
$^{75}$K$_{\mathrm{\mathrm{c}}}$ increases with decreasing
temperature presenting a strong anisotropy. At high temperature
the anisotropy is small where as with decreasing the temperature
the anisotropy is enhanced considerably.  If we compare
$^{75}$K$_{\mathrm{ab}}$, $^{75}$K$_{\mathrm{c}}$ at 50 K, it is
seen that anisotropy of the transferred field is really important
here and $^{75}$K$_{\mathrm{ab}}$ is 2.5 times larger than
$^{75}$K$_{\mathrm{c}}$.
\\
\indent From  Fig. {\ref{fig:Shift} it is seen that
$^{75}$K$_{\mathrm{ab}}$, $^{75}$K$_{\mathrm{c}}$ and
$^{75}$K$_{\mathrm{iso}}$ increases with decreasing  temperature
following the bulk susceptibility down to the temperature 50 K.
However, with further lowering the temperature down to 30 K shift
is decreasing with temperature leaving a maxima at around 50 K.
This maxima traced back the results of susceptibility and
specific heat study. Which indicates the ferromagnetic Co
ordering take place at $T_{\mathrm{c}}$ 75 K.  It is worthwhile
to mention that the bulk susceptibility is constantly increasing
with decreasing the temperature. It is well established that NMR
shift probes the local susceptibilty, therefore it is normally
not influenced by the small amount of the impurity. There are two
possibilities for the decrease of the shift. First, the system
have small amount of impurities which is not tracked by the XRD
measurement and makes the increase of susceptibility. Second,
there are polarisation effects on the Ce ions by the internal
magnetic field of the Co magnetism. Which in turn changes the
transferred hyperfine field below 50 K. The former one is
unlikely  because of the well-matched magnetisation result of
CeCoPO and LaCoAsO \cite{H. Ohta and K. Yoshimura-Phys rev.
B-2009}\cite{C. Krellner-Physica B-2009}. Therefore, the later
scenario is more likely. Below 50 K the decrement of the shift
reveals that the ferromagnetically ordered Co moment polarises
the Ce moment. Which eventually makes the magnetic structure
complicated and changes the hyperfine field in this system.
\begin{figure}
\includegraphics[scale=0.90]{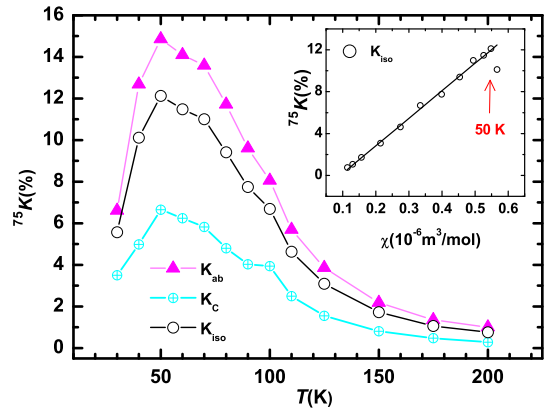}
\caption{\label{fig:Shift} Temperature dependence of the
$^{75}$K$_{ab}$, $^{75}$K$_{c}$ and $^{75}$K$_{iso}$. Inset shows
the plots of $^{75}$K$_{iso}$ against the bulk susceptibility
$\chi$ measured at 5 T. }
\end{figure}
\indent NMR probes the magnetism on a microscopic scale. As NMR
gives the local hyperfine field arising from the 4f-Ce and the
3d-Co ions. Usually for the itinerant 3d ions the negative core
polarisation is the dominant exchange mechanism whereas for
localised Ce moments the strong conduction electron polarization
contribution (Fermi contact interaction) becomes important.
Sometimes both fields cancel out each other leading to $K=0$
condition but often the Fermi contact interaction is more than one
order of magnitude larger \cite{Carter-Bennett-Kahan-1977}.
Therefore the total shift could be composed as $K=K_{4f}+K_{3d}$.
Furthermore $K_{4f}$ couples strongly on the effective
Ce-4f-moment. The effective moment is reduced because of CEF
splitting which results in a reduction of $K_{4f}$. This might
explain the shift maximum observed.
\\
\indent For an estimation of the hyperfine coupling constant
$^{75}$K$_{\mathrm{iso}}$ is plotted as a function of the bulk
susceptibility $\chi$ in the inset of Fig. {\ref{fig:Shift}. For
this we have used the susceptibility measured at 5T. We assume
that $\chi =\chi_{\mathrm{iso}}$, meaning that there is no
alignment or the texture in the CeCoAsO sample. From the inset it
is seen that $^{75}$K$_{\mathrm{iso}}$ is nicely following the
bulk susceptibility in the temperature range 200-50 K. From the
linear curve we have estimated the hyperfine coupling constants
at the $^{75}$As site, $^{75}A_{\mathrm{iso}}\approx $18
kOe/$\mu_{\mathrm{B}}$. The estimated $^{31}A_{\mathrm{iso}}$ for
the CeCoPO and LaCoAsO is around 14 kOe/$\mu_{\mathrm{B}}$ and
24.8 kOe/$\mu_{\mathrm{B}}$, respectively \cite{H. Ohta and K.
Yoshimura-Condmat-LaCoAsO-NMR-2009}\cite{Eva-CeCoPO-tesis-2010}.
Therefore for CeCoAsO the value of hyperfine coupling constants
is higher than that of CeCoPO system which could be interpreted
as being due to a weaker 3d-4f polarisation in CeCoAsO.
\\
\indent Below 15 K the additional line broadening shows up in the
spectra. Such a broadening can not be explained by the impurities
or disorder. As a first approach this line broadening traced back
the specific heat anomaly at 6.5 K and increase of $C/T$ below
1.8 K. This broadening is also typical for the onset of
correlation. However for this system at present with the
available data it is not settled whether this is the result of
complicated magnetic structure or it is due to the correlation.
For another 4f-3d pnictide NdCoAsO very recently McGurie et. al.
proposed from neutron scattering multiple phase transitions at
low temperature\cite{David Mandrus-NdCoAsO-condmat-2009}.
\section{Discussion and Summary}
Our findings on CeCoAsO yield several interesting phenomena which
shall now be discussed. The presented results point to a FM
ordering of Co ions at $T_{\mathrm{C}}$ =75 K similar to FM order
in the P-homolouge\cite{C. Krellner-Physica B-2009}. Similar to
CeCoPO unusual, but more S-like shaped $\chi(T)$ curves below
$T_{C}$ are observed. This is in contrast to LaCoPO and LaCoAsO
where $\chi(T)$ behave like textbook ferromagnets. For CeCoPO and
CeCoAsO a complex interplay of Co-3d moments with more localised
Ce-4f moments has to be considered. More into detail: Co 3d ions
order ferromagnetically at $T_{\mathrm{C}}$=75 K and the
resulting internal field transferred to the Ce site partially
polarizes the Ce-4f ions. Due to thermal excitation the
polarization is getting stronger towards lower temperatures
leading to an S-like shape of $\chi(T)$. Moreover, in the specific
heat an additional broad anomaly at around 6.5 K shows up, which
is different to CeCoPO system. This low temperature Schottky
anomaly indicates a complex level splitting of the ground state
of the rare earth moment by the internal field.
\\
\indent In order to get a deeper microscopic insight of this
system we have performed $^{75}$As NMR investigations. The NMR
shift is increasing with decreasing the temperature following the
susceptibility down to 50 K. The shift is decreased with lowering
the temperature further resulting in a broad maxima at 50 K. This
indicates that there is a change of hyperfine field below 50 K.
Furthermore, in the $^{75}$As NMR spectra the additional line
broadening below 15 K, traced back the specific heat anomaly at
6.5 K. Such broadening effect are typical for the onset of
correlations. But also a reconstruction of hyperfine fields
because of CEF splitting could be a possible explanation. To
fully understand the low temperature magnetism it deserves
further investigation specially with microscopic tool, for
instance neutron scattering and/or $\mu$SR.
\\
\indent Furthermore based on the presented results doping studies
on CeCoAsO might be fruitful in the context of superconductivity.
The superconducting state of the doped RFeAsO systems are
suggested to be unconventional nature. One key point to get the
superconductuvity in the CeFeAsO system is to suppress the Fe
magnetism by changing the carrier concentration. One approach
would be to substitute F in place of O, or substitute As by P or
Fe by small concentration of Co. In all cases for the specific
doping concentration one would get superconductivity
\cite{Zhao-JPCM-CeFe1-xCoxAsO-2010}\cite{Anton-CeFeAsPO-to be
published-2009}\cite{Chen-Li-2008}. Therefore still it is not
settled the nature of the carrier concentration required for this
CeFeAsO based superconductor. Apart from the Fe-based pnictides,
superconductivity was also found in LaNiPO and LaNiAsO.
Furthermore in the 122 relative Ba(Fe, Co)$_{2}$As$_{2}$
$T_{C}$`c up to  22 K are found. $^{59}$Co NMR investigations
clearly reveals that Co is
nonmagnetic\cite{Sefat-Prl-Ba(Fe1-xCox)2As-2010}\cite{Ning-Prb-Ba(Fe1-xCox)2As-2010}.
The absence of superconductivity in CeCo(As/P)O is not surprising
considering the fact that here Co carries a moment and long range
order is observed. Furthermore in these system there is a complex
interplay between the Ce-4f and 3d magnetism playing a crucial
role to control the magnetism. It is worthy to mention, as far as
3d magnetism is concern, there is a major difference between the
CeCoAsO to that of CeFeAsO. For CeFeAsO, Fe magnetism can be
tuned easily replacing As by P and SDW type transition
diminished, whereas for CeCoAsO, Co magnetism stays rigid with
that and SDW transition is absent. However still there is a
possibility to suppress the Co magnetism either by doping or
pressure, because an ordering temperature of around 70 K is
rather low for Co ordering. Therefore further research has to
answer the question whether the superconductivity appears in this
system after the suppression of the Co- magnetism.  Which will
eventually opens up the opportunity to understand the nature of
coupling between the 4f and itinerant electrons in the RTPnO
systems. And NMR/NQR would be the valuable tool to probe the
magnetism and superconductivity.
\\
\indent In summary we presented magnetisation, specific heat and
$^{75}$As NMR investigations on polycrystalline CeCoAsO. The
magnetisation and specific heat data reveal that in this system
Co orders ferromagnetically at 75 K. Moreover specific heat study
shows an Schottky anomaly at low temperature at around 6.5 K,
which is likely due to the level splitting of the ground state of
rare earth moment by the internal field. Furthermore the analysis
of the $^{75}$As NMR spectra clearly demonstrate the strong shift
anisotropy towards lower temperature. Moreover the breakdown of
the $K$ vs. $\chi$ linearity below 50 K might be the signature of
CEF interaction.


\bibliography{apssamp}

\end{document}